\documentclass[11pt]{article}
\usepackage[english]{babel}
\usepackage[dvips]{graphics,color}
\usepackage{newlfont,epsfig}
\usepackage{graphicx}
\begin{document}
\date{}
\begin{center}
{\huge{Dynamics of two atoms coupled to a cavity field}}
\end{center}
\vspace{0.3cm}
\begin{center}
{\large J.A. Roversi$^{1}$, A. Vidiella-Barranco$^{1}$ and H. Moya-Cessa$^{2}$}
\end{center}
\vspace{0.2cm}
\begin{center}
{\it\normalsize{$^{1}$Instituto de F\'\i sica ``Gleb Wataghin'' - Universidade Estadual
de Campinas}}\\
{\it\normalsize{ 13083-970   Campinas  SP  Brazil}}\\
{\it\normalsize{$^{2}${INAOE, Coordinaci\'on de Optica, Apdo. Postal 51 y 216, 72000 Puebla, Puebla, 
Mexico}}}\\
\end{center}
\vspace{0.2cm}
\begin{abstract}
We investigate the interaction of two two-level atoms with a single mode cavity
field. One of the atoms is exactly at resonance with the field, while the other is
well far from resonance and hence is treated in the dispersive limit. We find that 
the presence of the non-resonant atom produces a shift in the Rabi frequency of the 
resonant atom, as if it was detuned from the field. We focus on the discussion of the 
evolution of the state purity of each atom.%
\end{abstract}

\section{Introduction}

Two atoms in interaction with a cavity field constitutes an interesting 
quantum optical system \cite{gontier,guo}. It involves three fully quantized subsystems 
experimentally accesible and which may be handled (from the theoretical point of view)
without too many approximations. The investigation of tripartite systems is particularly 
attractive specially due to the interest in three-system entanglement with implications on 
quantum information, for instance \cite{pati}. We have two subsystems (atoms) in 
principle able to {\it store} quantum information and one 
subsystem (field) able to {\it transmit} quantum information, and one may ask about what
kind of correlations could arise between the atomic subsystems in virtue of their
interaction with the field. Nevertheless, it seems that such a physical situation has 
been not fully explored, and deserves further investigation. In most treatments, 
the atoms are supposed to be either both resonant with the electromagnetic field 
\cite{gontier} or both nonresonant \cite{guo}. Here we consider a different 
(asymmetric) situation; one atom resonant with the cavity field and the other very far 
from resonance. The atoms will basically interact via the cavity field, and will show a
modified dynamics due to the presence of the other atom. In Section 2 we present the 
solution of our model, and in Section 3 we discuss a few aspects of the reduced dynamics of 
the atomic subsystems. In Section 4 we summarize our conclusions.

\section{Model}

We consider that two (two-level) atoms are coupled to the same mode of the 
electromagnetic field inside a high-Q cavity, so that in the present approach 
losses will be neglected. We are also not considering direct coupling (dipole-dipole type, 
for instance) between the atoms in the present work. Atom 1 is in resonance with the cavity field (of frequency $\omega$), having a 
coupling constant $\lambda_1$ while atom 2 is out of resonance (atomic frequency 
$\omega_2$). We may therefore treat the atom 2-field interaction in the so-called 
dispersive limit \cite{luis}, which results in the effective interaction hamiltonian 
$\hat{H}^{(2)}_i=\lambda_2 \hat{a}^\dagger\hat{a} \sigma_z^{(2)}$.
The corresponding complete hamiltonian may be written as
\begin{equation}
\hat{H}=\hbar\Big[ \omega\hat{n} + \omega\sigma_z^{(1)} 
+ \omega_2\sigma_z^{(2)} + \lambda_1\big(\hat{a}^\dagger\sigma_-^{(1)} 
+\sigma_+^{(1)}\hat{a}\big)  
+ \lambda_2 \hat{n} \sigma_z^{(2)} \big].\label{firsthamil}
\end{equation}
The effective atom 2-field coupling constant is $\lambda_2=\lambda/\delta$, where $\lambda$
is the dipole coupling constant, and $\delta=\omega_2-\omega$. The dispersive approximation
is valid provided that $| \delta |\gg \sqrt{n+1}\, \lambda$. We are also considering that
atom 2 is weakly coupled to the field in the sense that $\lambda_2 \ll \lambda_1$. 
The atomic operator $\sigma_j^{(i)}$ refers to the $i$-th atom and 
$\hat{n}=\hat{a}^\dagger\hat{a}$ is the photon number operator.
 
The hamiltonian $\hat{H}$ above may be further transformed into the more convenient
interaction picture hamiltonian
\begin{equation}
\hat{H}_I = \hat{U}_0\hat{H}\hat{U}_0^\dagger=(\delta+\lambda_2\hat{n})\sigma_z^{(2)} +
\lambda_1\left(\hat{a}^\dagger\sigma_-^{(1)} +\sigma_+^{(1)}\hat{a}\right),
\end{equation}
where
\begin{equation}
\hat{U}_0(t)=\exp[-i\omega t(\hat{n}+\sigma_z^{(1)}+\sigma_z^{(2)})].
\end{equation}

We may assume that initially the atoms are decoupled from the field, so that we
have an initial product state for the whole system
\begin{equation}
|\Psi(0)\rangle=|\varphi\rangle \otimes (a_1|g_1\rangle + b_1|e_1\rangle) \otimes
(a_2|g_2\rangle + b_2|e_2\rangle),
\end{equation}
i.e., both atoms prepared in superpositions of their energy
eigenstates, and with $|\varphi\rangle=\sum A_n |n\rangle$ being the initial 
arbitrary (pure) state of the field. After solving Schr\"odinger's equation, 
we obtain the following time dependent state vector
\begin{eqnarray}
|\Psi(t)\rangle&=&\sum_{n=0}^\infty \big\{C_{1;n} 
e^{-i[\delta+\lambda_2(n-1/2)]t} |n-1\rangle |e_1\rangle |e_2\rangle
+ C_{2;n} e^{-i[\delta+\lambda_2(n-1/2)]t}
|n\rangle |g_1\rangle |e_2\rangle\nonumber \\ 
&+& C_{3;n} 
e^{-i(\delta_n+\lambda_2/2)t} |n\rangle |e_1\rangle |g_2\rangle
+ C_{4;n}  
e^{-i[\delta+\lambda_2(n+1/2)]t} |n+1\rangle |g_1\rangle |g_2\rangle\big\},
\end{eqnarray}

with the following coefficients
\begin{eqnarray}
C_{1;n}&=&\left(\cos\Delta_nt +
i\frac{\lambda_2}{2}\sin\frac{\Delta_nt}{\Delta_n}\right)A_{n-1}b_1 b_2 
-i\lambda_1\sqrt{n}\sin\frac{\Delta_nt}{\Delta_n}A_{n}a_1 b_2 \\ \nonumber
C_{2;n}&=& \left( \cos\Delta_nt - i\frac{\lambda_2}{2}\sin\frac{\Delta_nt}
{\Delta_n}\right) A_{n}a_1 b_2 + i\lambda_1\sqrt{n}\sin\frac{\Delta_nt}{\Delta_n}A_{n-1}
b_1 b_2  \\ \nonumber
C_{3;n}&=&\left(\cos\Delta_{n+1}t -
i\frac{\lambda_2}{2}\sin\frac{\Delta_{n+1}t}{\Delta_{n+1}}\right)A_{n}b_1
a_2 - i\lambda_1\sqrt{n+1}\sin\frac{\Delta_{n+1}t}{\Delta_{n+1}}A_{n+1}
a_1 a_2 \\ \nonumber
C_{4;n}&=& \left( \cos\Delta_{n+1}t + i\frac{\lambda_2}{2}\sin
\frac{\Delta_{n+1}t}{\Delta_{n+1}}\right) A_{n+1}a_1 a_2 - i\lambda_1\sqrt{n+1}\sin\frac{\Delta_{n+1}t}{\Delta_{n+1}}A_{n}
b_1 a_2 
\end{eqnarray}
where 
\begin{equation}
\Delta_n=\sqrt{\frac{\lambda_2^2}{4} + n\lambda_1^2}.
\end{equation} 
This means that one of the effects of atom 2 is to introduce a phase shift of $\lambda_2^2/4$ in such a way that 
we end up with ``modified'' generalized Rabi frequencies $\Delta_n$. 

\section{Reduced dynamics}

As an example we may consider a simple initial conditional for the global state (two atoms $+$ field )
\begin{equation}
|\Psi(0)\rangle=|0\rangle\otimes|e_1\rangle)\otimes(a_2|g_2\rangle + b_2|e_2\rangle),
\end{equation}
or the field in the vacuum state, atom 1 in the excited state and atom 2 in a
superposition state. After tracing over the atom 2 as well as the field variables,
we obtain the following reduced density operator for atom 1
\begin{equation}
\hat{\rho}_1(t)=\Big( \cos^2\Delta_1 t + \frac{\lambda_2^2}{4}
\frac{\sin^2\Delta_1 t}{\Delta_1^2} \Big) |e_1\rangle\langle e_1 | 
+\lambda_1^2\frac{\sin^2\Delta_1t}{\Delta_1^2}\, |g_1\rangle\langle g_1 |,
\end{equation}
with $\Delta_1=\sqrt{\lambda_1^2 + \lambda_2^2/4}$.
We note that an important effect of
the non-resonant atom 2 on the resonant one (atom 1) is to introduce a frequency 
shift of $\lambda_2/2$, which is equivalent to a (atom 1-cavity) detuning, 
although atom 1 is actually in resonance with the cavity field. Such a result may be viewed as a consequence of
the fact that the presence of the nonresonant atom 2 causes a ``cavity frequency pulling'' i.e., a shift in the cavity frequency \cite{haro}.
The coupling with atom 2 
also introduces the oscillatory extra term $\lambda_2^2\,\sin^2\Delta_1/4\Delta_1^2$ 
in the coefficient of $|e_1\rangle\langle e_1 |$.
The reduced density operator for atom 2 will be
\begin{eqnarray}
\hat{\rho}_2(t)&=& |a_2|^2\, |g_2\rangle\langle g_2| + |b_2|^2\, |e_2\rangle\langle e_2|\\ \nonumber
&+&\Big[\Big( \cos \Delta_1 t - i \frac{\lambda_2}{2}
\frac{\sin\Delta_1t}{\Delta_1} \Big)^2  
+\lambda_1^2\frac{\sin^2\Delta_1t}{\Delta_1^2} \Big] e^{2i(\delta+\lambda_2/2)}
a_2b^*_2\,|g_2\rangle\langle e_2 | \\ \nonumber
&+&\Big[\Big( \cos\Delta_1 t + i \frac{\lambda_2}{2}
\frac{\sin\Delta_1 t}{\Delta_1} \Big)^2 
+\lambda_1^2\frac{\sin^2\Delta_1t}{\Delta_1^2} \Big] e^{-2i(\delta+\lambda_2/2)}
a^*_2b_2\,|e_2\rangle\langle g_2 |
\end{eqnarray}
and finally for the cavity field
\begin{equation}
\hat{\rho}_f(t)=\Big( \cos^2\Delta_1 t + \frac{\lambda_2^2}{4}
\frac{\sin^2\Delta_1 t}{\Delta_1^2} \Big) |0\rangle\langle 0 | 
+\lambda_1^2\frac{\sin^2\Delta_1t}{\Delta_1^2}\, |1\rangle\langle 1 |.
\end{equation}

We would like to remark that even when atom 2 is initially prepared in either 
$|g_2\rangle$ $(b_2=0)$ or $|e_2\rangle$ $(a_2=0)$, which are stationary states for atom 2,
there will be substantial changes to the evolution of atom 1 due to phase shifts 
introduced in the field through the dispersive interaction. The subsystem atom 2 plays the 
role of a ``single particle reservoir'', in the sense that it will induce modifications 
in the subsystem atom 1 without having its state changed. Therefore the presence of nearby atoms,  
out of resonance with the field, may influence the dynamics of a resonant atom in a way 
that they might become a source of decoherence. If atom 2 is initially prepared
in a superposition of states $|g_2\rangle$ and $|e_2\rangle$, we clearly see the
influence of atom 2 upon the dynamics of atom 1, given by terms such as
$\lambda_1^2\sin^2\Delta_1t/\Delta_1^2$, which are part of the coefficients of both
$|g_2\rangle\langle e_2 |$ and $|e_2\rangle\langle g_2 |$.
In order to better characterize the subsystems' evolution we show 
some plots of the state purity, defined as $\zeta_i=1-Tr\hat{\rho}_i^2$. We have,
for atom 1
\begin{equation}
{\zeta}_1(t)=2\, 
\frac{\sin^2\left(\sqrt{1+\Theta^2}\lambda_1t\right)}{1 + \Theta^2}
\left[1-\frac{\sin^2\left(\sqrt{1+\Theta^2}\lambda_1t\right)}{1 + \Theta^2}\right],
\end{equation}
and for atom 2
\begin{equation}
{\zeta}_2(t)=8\, a_2^2\, b_2^2\, \Theta^2 \frac{\sin^4\left(\sqrt{1+\Theta^2}\lambda_1t\right)}
{\left(1 + \Theta^2 \right)^2},
\end{equation}
with $\Theta= \lambda_2/2\lambda_1$.

In Fig. 1 we have the plot of the purities relative to the states of atom 1, $\zeta_1$,
and atom 2, $\zeta_2$,  as a function of the scaled time $\lambda_1 t$ (for $\lambda_1=1.0$ and
$\lambda_2=0.2$). We note a little deviation from ordinary Rabi oscillations, due to the presence of  atom 
2. For comparison purposes, we have chosen atom 2 initially in a superposition state, or $a_2=b_2=1/\sqrt{2}$, so that $\zeta_2\neq 0$. If we increase the value of $\lambda_2$ to $\lambda_2=0.5$ for instance, we have the situation shown in Fig. 2. We note a stronger modulation in the oscillations and a clear departure from ordinary 
Rabi oscillations is verified as atom 2 becomes more strongly coupled to the field. Further studies about such
a tripartite system are being carried out and will be presented elsewhere.

\section{Conclusions}

We have investigated the problem of the interaction of two two-level atoms with a mode of the quantized field. 
We have considered one of the atoms (atom 1) to be in exact resonance with the field while the other (atom 2) is 
very far from resonance. 
We have found that, even in the case that atom 2 remains in its initial state, important changes will occur in the dynamics of atom 1. The phase changes induced by the dispersive coupling of atom 2 to the field causes a shift in the Rabi frequency, as well as a modulation in the oscillations of the atomic 
inversion.

\begin{figure}[th]
\centerline{\psfig{file=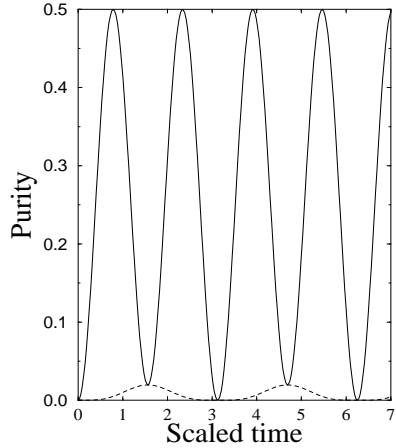,height=5cm,width=4cm}}
\vspace*{8pt}
\caption{State purity $\zeta$ as a function of the scaled time $\lambda_1 t$ relative to atom 1 (solid line) and 
atom 2 (dashed line), for $\lambda_1=1$ and $\lambda_2=0.2$.}
\end{figure}

\begin{figure}[th]
\centerline{\psfig{file=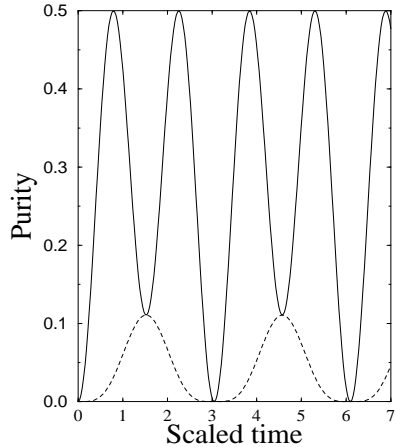,height=5cm,width=4cm}}
\vspace*{8pt}
\caption{State purity $\zeta$ as a function of the scaled time $\lambda_1 t$ relative to atom 1 (solid line) and atom 2 (dashed line) for $\lambda_1=1$ and $\lambda_2=0.5$.}
\end{figure}

\section*{Acknowledgements}

This work was partially supported by CNPq (Conselho Nacional para o 
Desenvolvimento Cient\'\i fico e Tecnol\'ogico, Brazil), CONACyT (Consejo 
Nacional de Ciencia y Tecnolog\'\i a, M\'exico)and it is linked 
to the Optics and Photonics Research Center (FAPESP, Brazil).

\end{document}